\documentclass[aps,prb,twocolumn,showpacs,superscriptaddress]{revtex4-1}

\usepackage{graphicx}  % needed for figures
\usepackage{dcolumn}   % needed for some tables
\usepackage{bm}        % for math
\usepackage{amssymb}   % for math
\usepackage{subcaption}
\usepackage{amsmath}

\begin{document}

\preprint{APS/123-QED}

\title{Neutron Diffraction and $\mu$SR Studies of Two Polymorphs of Nickel Niobate (NiNb$_2$O$_6$)}% Force line breaks with \\

\author{T.J.S. Munsie}
\affiliation{Department of Physics and Astronomy, McMaster University, Hamilton, Ontario L8S 4M1, Canada.}

\author{M.N. Wilson}
\affiliation{Department of Physics and Astronomy, McMaster University, Hamilton, Ontario L8S 4M1, Canada.}

\author{A. Millington}
\affiliation{Department of Physics and Astronomy, McMaster University, Hamilton, Ontario L8S 4M1, Canada.}

\author{C.M. Thompson}
\affiliation{Department of Chemistry, Purdue University, West Lafayette, Indiana 47907-2084, USA}

\author{R. Flacau}
\affiliation{Canadian Neutron Beam Centre, Chalk River, Ontario K0J 1J0, Canada}

\author{C. Ding}
\affiliation{Department of Physics, Zhejiang University, Hangzhou 310027, China.}

\author{S. Guo}
\affiliation{Department of Physics, Zhejiang University, Hangzhou 310027, China.}

\author{Z. Gong}
\affiliation{Department of Physics, Columbia University, New York, NY 10027, USA.}

\author{A.A. Aczel}
\affiliation{Quantum Condensed Matter Division, Neutron Sciences Directorate,
Oak Ridge National Laboratory, Oak Ridge, TN 37831, USA}

\author{H.B. Cao}
\affiliation{Quantum Condensed Matter Division, Neutron Sciences Directorate,
Oak Ridge National Laboratory, Oak Ridge, TN 37831, USA}

\author{T.J. Williams}
\affiliation{Quantum Condensed Matter Division, Neutron Sciences Directorate,
Oak Ridge National Laboratory, Oak Ridge, TN 37831, USA}

\author{H.A. Dabkowska}
\affiliation{Brockhouse Institute for Materials Research, McMaster University, Hamilton, Ontario L8S 4M1, Canada.}

\author{F. Ning}
\affiliation{Department of Physics, Zhejiang University, Hangzhou 310027, China.}

\author{J.E. Greedan}
\affiliation{Department of Chemistry, McMaster University, Hamilton, Ontario L8S 4M1, Canada.}
\affiliation{Brockhouse Institute for Materials Research, McMaster University, Hamilton, Ontario L8S 4M1, Canada.}

\author{G.M. Luke}
\affiliation{Department of Physics and Astronomy, McMaster University, Hamilton, Ontario L8S 4M1, Canada.}
 \affiliation{Brockhouse Institute for Materials Research, McMaster University, Hamilton, Ontario L8S 4M1, Canada.}
 \affiliation{Canadian Institute for Advanced Research, Toronto, Ontario M5G 1M1, Canada.}
\date{\today}% It is always \today, today,

\begin{abstract}
Neutron diffraction and muon spin relaxation ($\mu$SR) studies are presented for the newly characterized polymorph of NiNb$_2$O$_6$ ($\beta$-NiNb$_2$O$_6$) with space group P4$_2$/n and $\mu$SR data only for the previously known columbite structure polymorph with space group Pbcn. The magnetic structure of the P4$_2$/n form was determined from neutron diffraction using both powder and single crystal data. Powder neutron diffraction determined an ordering wave vector $\vec{k}$ = ($\frac{1}{2},\frac{1}{2},\frac{1}{2}$). Single crystal data confirmed the same $\vec{k}$-vector and showed that the correct magnetic structure consists of antiferromagnetically-coupled chains running along the a or b-axes in adjacent Ni$^{2+}$ layers perpendicular to the c-axis, which is consistent with the expected exchange interaction hierarchy in this system. The refined magnetic structure is compared with the known magnetic structures of the closely related tri-rutile phases, NiSb$_2$O$_6$ and NiTa$_2$O$_6$. $\mu$SR data finds a transition temperature of $T_N$~$\sim$~15 K for this system, while the columbite polymorph exhibits a lower $T_N$~$=$~5.7(3) K. Our $\mu$SR measurements also allowed us to estimate the critical exponent of the order parameter $\beta$ for each polymorph. We found $\beta$~$=$~0.25(3) and 0.16(2) for the $\beta$ and columbite polymorphs respectively. The single crystal neutron scattering data gives a value for the critical exponent $\beta$~$=$~0.28(3) for $\beta$-NiNb$_2$O$_6$, in agreement with the $\mu$SR value. While both systems have $\beta$ values less than 0.3, which is indicative of reduced dimensionality, this effect appears to be much stronger for the columbite system. In other words, although both systems appear to well-described by $S$~$=$~1 spin chains, the interchain interactions in the $\beta$-polymorph are likely much larger.

%Single crystal data showed that the correct structure is described by a combination of basis vectors belonging to the $\Gamma3 + \Gamma7$ irreducible representations and consists of antiferromagnetically coupled chains oriented along either the a or b axes normal to c. This structure is compared with the known magnetic structures of the closely related tri-rutile phases, NiSb$_2$O$_6$ and NiTa$_2$O$_6$. Neutron diffraction and $\mu$SR data confirm the critical temperature of 13.6(0.2) and 14.5(0.3) K, respectively, for the P$4_2/n$ form with an estimate of the critical exponent β = 0.28(0.03) and 0.25(0.02), respectively, and agreeing within error. For the Pbcn polymorph the behaviour of the order parameter is measured for the first time using $\mu$SR, resulting in T$_N$ = 5.7(0.3)K, in good agreement with previous work, and β = 0.16(0.02), a remarkably low value.

\end{abstract}

\maketitle

%\tableofcontents

\section{Introduction}
\begin{figure*}
	\begin{subfigure}[]{.5\textwidth}
		\centering
		\includegraphics[width=.7\linewidth]{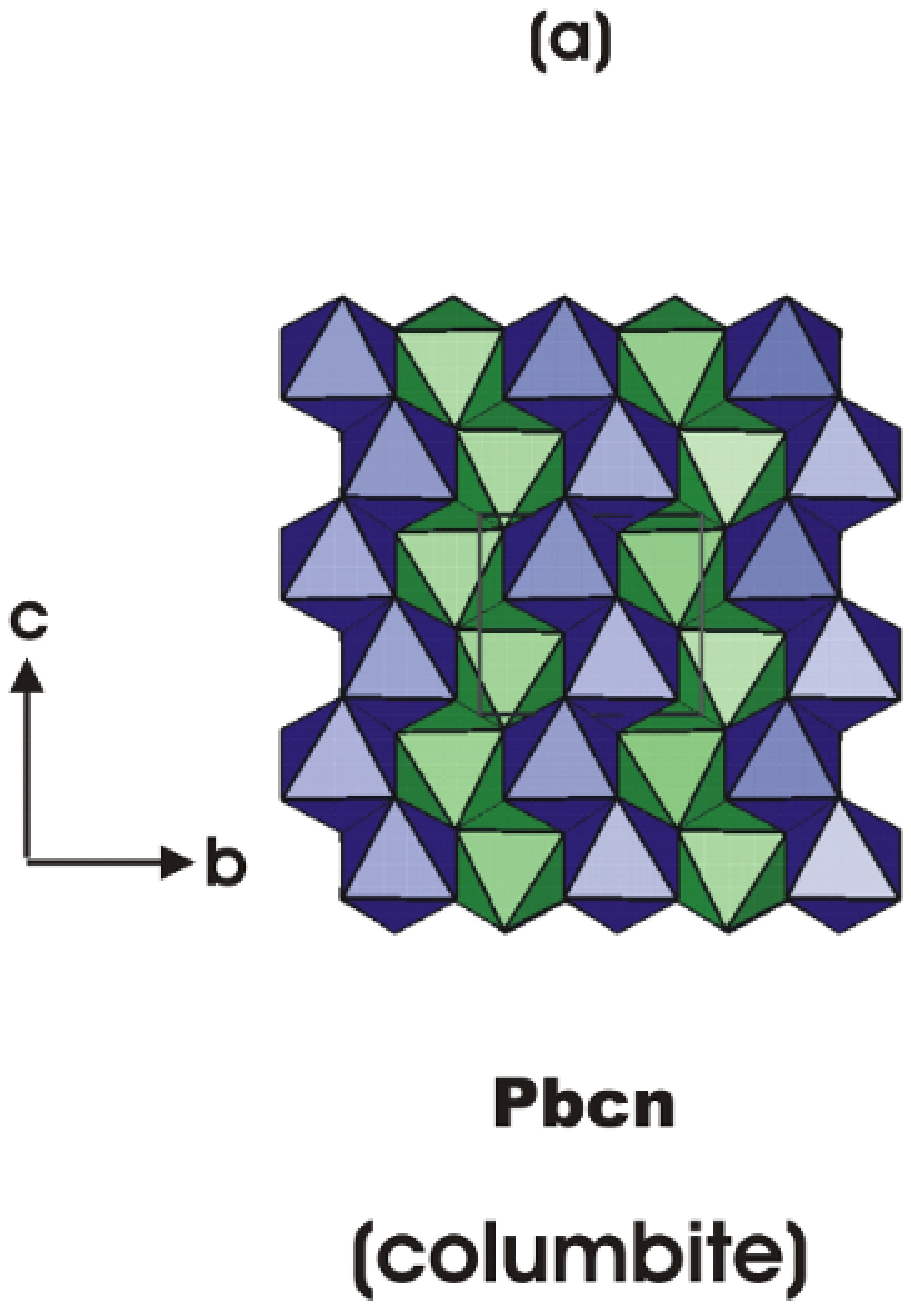}
		\caption{}
		\label{fig:VESTAPbcn}
	\end{subfigure}%
	\begin{subfigure}[]{.5\textwidth}
		\centering
		\caption{}
		\includegraphics[width=.7\linewidth]{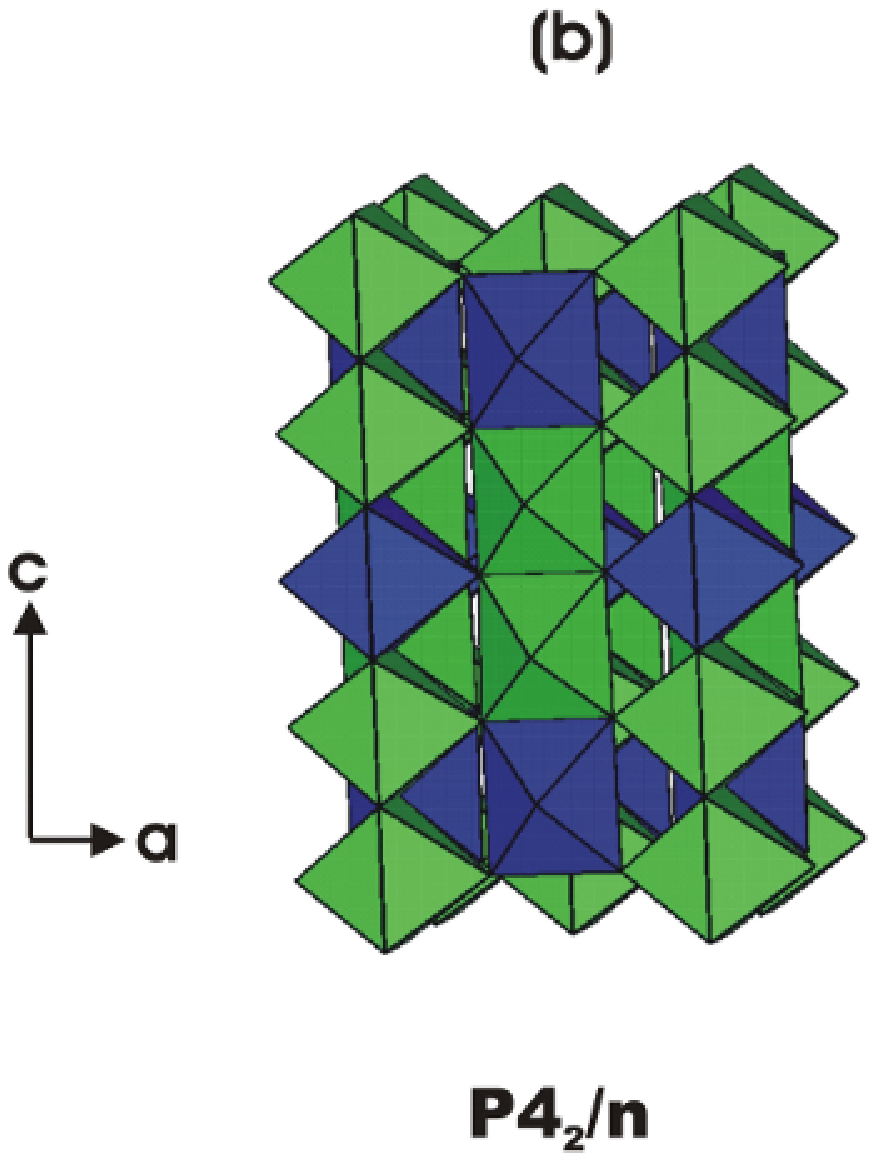}
		\label{fig:VestaSG86}
	\end{subfigure}
	\caption{(Colour online)Two polymorphs of NiNb$_2$O$_6$. The blue (dark) octahedra contain nickel, the green (light) octahedra contain niobium. Images created with VESTA~\cite{momma2011}. (a) A bc plane view of the columbite structure showing the zigzag chains of edge sharing Ni-O octahedra along c. Data from ~\cite{weitzel1976}. (b) An ac plane view of the P4$_2$/n polymorph showing the absence of the edge-sharing Ni-O octahedra.}
	\label{fig:VestaFigs}
\end{figure*}

Most transition metal niobates, ANb$_2$O$_6$, crystallize in the columbite space group (Pbcn, space group 60)~\cite{weitzel1976,sarvezuk2011,wichmann1983}, seen in Fig.~\ref{fig:VESTAPbcn}. In columbite, the zigzag edge-sharing chains of AO$_6$ octahedra are oriented parallel to the c-axis. These are separated by Nb-O edge-sharing chains in both the bc plane (shown) and by two such chains in the ac-plane (not shown) leading to dominant one dimensional magnetic interactions. The case of cobalt niobate (CoNb$_2$O$_6$) has been of strong, recent interest~\cite{kobayashi1999,kunimoto1999,coldea2010,morris2014,kinross2014,cabrera2014, liang2015} since the cobalt ion has an effective spin of $\frac{1}{2}$ and exhibits a quantum phase transition in a modest applied field, with clear experimental signatures of quantum critical phenomena previously predicted for the transverse field Ising model \cite{kinross2014}.

Another columbite structure niobate that has been studied and is of special interest is NiNb$_2$O$_6$, where Ni$^{2+}$ ($S$~$=$~1) replaces Co$^{2+}$ (S$_{eff}$ = $\frac{1}{2}$). Short range ferromagnetic spin correlations develop in this material on cooling due to the dominant intrachain interactions, with long range antiferromagnetic order ultimately being achieved at T$_N$~$=$~5.7 K due to finite interchain interactions~\cite{heid1996, yaeger1977}. The magnetic structure of columbite NiNb$_2$O$_6$ was solved from powder neutron diffraction data~\cite{weitzel1976}, but the order parameter has not been reported.  

\begin{figure*}
	\begin{subfigure}[b]{.5\textwidth}
		\centering
		\includegraphics[width=.7\linewidth]{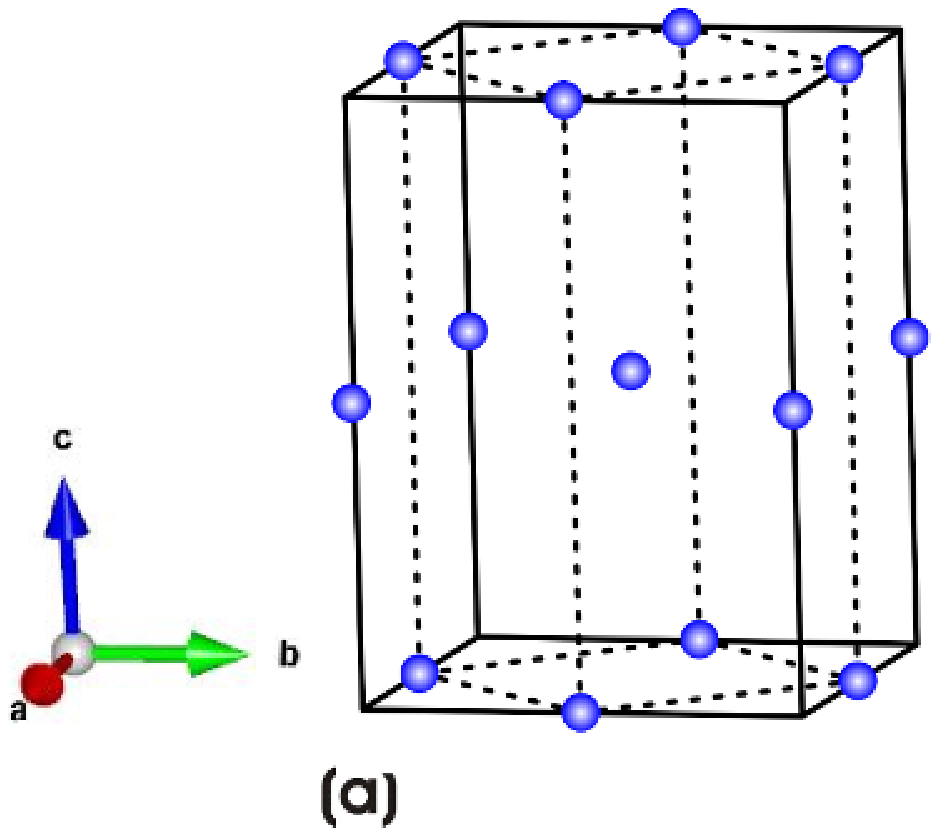}
		\caption{}
		\label{fig:NiSitesBeta}
	\end{subfigure}%
	\begin{subfigure}[b]{.5\textwidth}
		\centering
		\includegraphics[width=.65\linewidth]{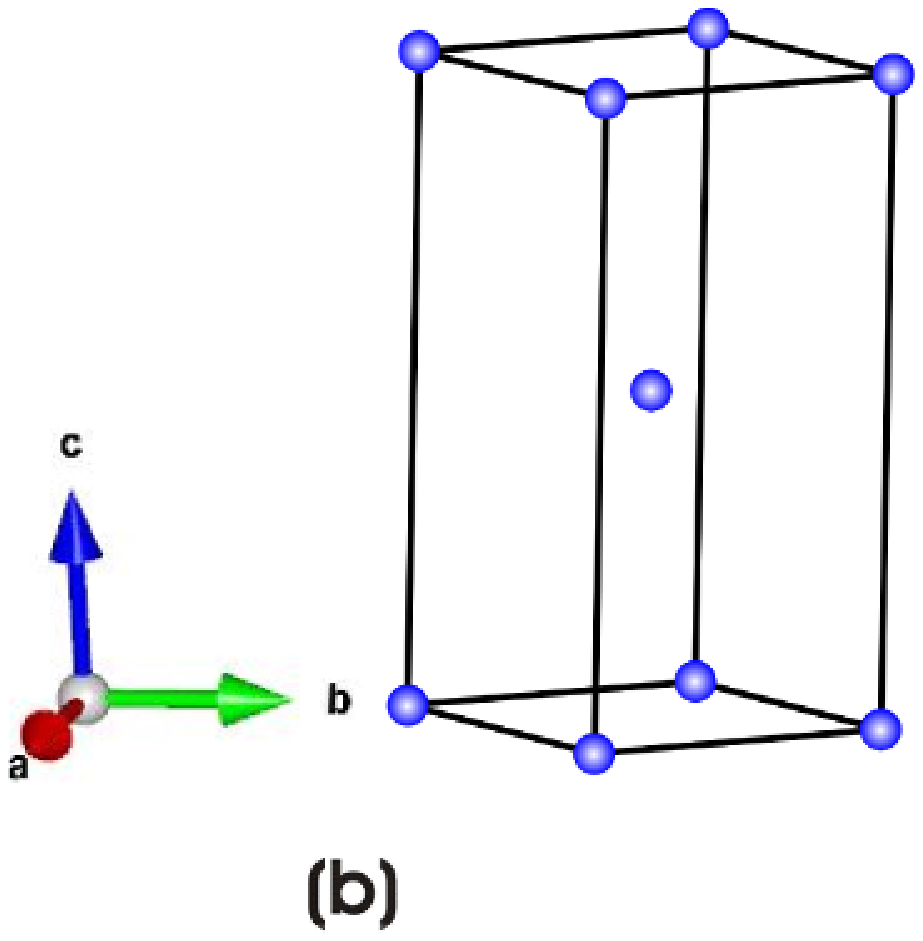}
		\caption{}
		\label{fig:NiSitesTriRutile}
	\end{subfigure}
	\caption{A comparison of the nickel sites in (a) the $\beta$ polymorph of nickel niobate and (b) the trirutile nickel tantalate. In Fig.~\ref{fig:NiSitesBeta} the cell of the same dimensions as the tantalate is outlined, demonstrating that for both materials the nickel sublattice has the same body-centred tetragonal symmetry.}
	\label{fig:NiSites}
\end{figure*}

%The aim of our research was to replace the cobalt atoms of CoNb$_2$O$_6$ with nickel atoms to enable a direct comparison of the magnetic properties for $S_{eff}$~$=$~1/2 and $S$~$=$~1 spin chain columbites. 
Initial attempts to synthesize NiNb$_2$O$_6$, following previously reported growth techniques~\cite{prabhakaran2003}, resulted in the formation of phases of both the expected columbite structure and a new polymorph crystallizing in P4$_2$/n (space group 86), a space group not previously reported in members of the AB$_2$O$_6$ family~\cite{beck2012}. Structural features of the new polymorph, $\beta$-NiNb$_2$O$_6$, are shown in Figs.~\ref{fig:VestaSG86} and \ref{fig:NiSitesBeta}. Of note is the absence of edge-sharing NiO$_6$ octahedra as found in the columbite form, shown in Fig.~(\ref{fig:VESTAPbcn}). Instead, the Ni$^{2+}$ ions are seen to form a body-centred tetragonal (b.c.t.) lattice, which is identical to that found in the closely related phases NiTa$_2$O$_6$ and NiSb$_2$O$_6$ that crystallize in the tri-rutile (TR) structure, P4$_2$/mnm (space group 136). P4$_2$/mnm and P4$_2$/n are related in a group-subgroup sense via the intermediate P4$_2$/m symmetry~\cite{hahn1983}. The principle difference between the structures in the P4$_2$/n group for nickel niobate and the more symmetric TR form (P4$_2$/mnm) is the position of the O$^{2-}$ ions which results in a highly-distorted NbO$_6$ octahedron with six different Nb-O distances in the former. This is attributed to a second order Jahn-Teller distortion~\cite{halasyamani1998}. As a result of these distortions, the P4$_2$/n unit cell has twice the volume of a TR cell with a = $\sqrt{2\text{a}_{TR}}$, c = c$_{TR}$.

Body-centred tetragonal sublattices are well-known to give rise to low dimensional antiferromagnetism; for example, K$_2$NiF$_4$ and many related materials exhibit strong two-dimensional spin correlations~\cite{deJongh2012}. Spin coupling in the third dimension is frustrated by the b.c.t. geometry. The cases of NiTa$_2$O$_6$ and NiSb$_2$O$_6$ are most relevant here. The bulk magnetic properties of these two systems are very similar to those of NiNb$_2$O$_6$, as summarized in Table~\ref{tbl:BulkMagProp}.

\begin{table}
	\renewcommand{\arraystretch}{1.2}
\begin{tabular}{|c|c|c|c|c|c|}
	\hline 
B	& $\mu_{eff}\left(\mu_B\right)$ & $\theta_c\left(K\right)$ & $T\left(\chi_{max}\right)\left(K\right)$ & $T_N\left(K\right)$ & Ref.\\ 
	\hline \hline
Ta	& 4.10 & -41 & 25.5 & 10.3 & ~\cite{takano1970, kremer1988}\\ 
	\hline 
Sb	& 3.00 & -50 & $\sim$ 30 & 2.5 & ~\cite{ramos1991,ramos1992} \\ 
	\hline 
Nb	& 3.35 & -37 & 22.5 & 14.0 & ~\cite{munsie2016}\\ 
	\hline 
\end{tabular} 
\caption{A summary of the bulk magnetic properties of NiB$_2$O$_6$ (B = Ta, Sb, Nb) family.}
\label{tbl:BulkMagProp}
\end{table}

\begin{figure}
	\centering
	\includegraphics[width=0.9\linewidth]{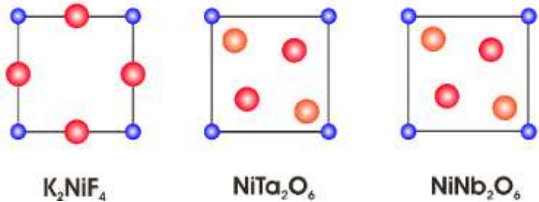}
	\caption{(Colour online) Ni (blue/small) and oxide ion (red and orange/large) positions in the K$_2$NiF$_4$, TR(NiTa$_2$O$_6$ and NiSb$_2$O$_6$) and NiNb$_2$O$_6$ structures shown along a (001) square plane defined by the Ni ions. For K$_2$NiF$_4$ the oxide ions and the Ni ions are in the same plane. In NiTa$_2$O$_6$ and NiNb$_2$O$_6$ one set of O ions (red) are either exactly in the plane of the Ni ions (Ta) or only slightly out of the plane by 0.096\AA\  (Nb), while the other set (orange) is either 1.55\AA\  (Ta) or 1.57\AA\  (Nb) from the plane.}
	\label{fig:spacegroupninb2o63}
\end{figure}

All three materials show a broad $\chi_{max}$ above 20K followed by long range order at significantly lower temperatures. Recent studies argue that NiTa$_2$O$_6$ and NiSb$_2$O$_6$ are best described as Heisenberg $S$~$=$~1 linear chain antiferromagnets~\cite{ramos1992,law2014}. Thus, the magnetic dimensionality is a strong function of the positions of the ligands with respect to the metal ions in the b.c.t. sublattice. This is illustrated in Fig.~\ref{fig:spacegroupninb2o63} for the K$_2$NiF$_4$, TR and the P4$_2$/n structures. In K$_2$NiF$_4$ the strongest exchange pathway involves nearest-neighbour Ni--O--Ni 180$^\circ$ linkages resulting in two-dimensional magnetism. For the TR structure the dominant exchange is between next nearest neighbours involving a M--O--O--M 180$^\circ$ pathway along [1 1 0], which gives rise to the one-dimensional magnetism. Recent Density Functional Theory (DFT) calculations show that the exchange constant for this pathway is $\sim$40 times stronger than any other in the TR structure~\cite{ehrenberg1998}. The ligand positions in P4$_2$/n are very similar to those in TR structure, with the main difference being that one set of O ions lies slightly out of the plane of the Ni ions, whereas in the TR structure these ions lie in the same plane. This is shown in Fig.~\ref{fig:spacegroupninb2o63}. 

The magnetic structures of the two TR phases are surprisingly different in spite of similar unit cell constants which differ by only 1-2\%. For NiTa$_2$O$_6$, the magnetic structure is quite complex with $\vec{k}$ = [$\frac{1}{4},\frac{-1}{4},\frac{1}{2}$]. The magnetic unit cell is orthorhombic with dimensions a$_M$ =  $\sqrt{2}$a, b$_M$ = $2\sqrt{2}$a and c$_M$ = 2c, V$_{M}^{cell}$ = 8 V$^{cell}$ ~\cite{law2014,ehrenberg1998}. On the other hand, for NiSb$_2$O$_6$ the propagation vector is $\vec{k}$ = [$\frac{1}{2}, 0, \frac{1}{2}$] and the magnetic unit cell dimensions (also orthorhombic) are a$_M$ = a, b$_M$ = 2b, c$_M$ = 2c, V$_M^{cell}$ = 4 V$^{cell}$ ~\cite{law2014}. Based on these findings, it is of particular interest to determine the magnetic structure for $\beta$-NiNb$_2$O$_6$. 

We collected neutron diffraction data for $\beta$-NiNb$_2$O$_6$ on a powder sample at the Canadian Neutron Bean Centre and on a single crystal sample at Oak Ridge National Laboratory at selected temperatures between 4 and 20 K. Using $\mu$SR we studied both polymorphs of nickel niobate in their powder form with zero applied field as a function of temperature to measure their order parameters and to study spin dynamics.  

\section{Experimental Methods}
A description of the growth of single crystal of both NiNb$_2$O$_6$ polymorphs has been reported~\cite{munsie2017a}.

Some single crystals of $\beta$-NiNb$_2$O$_6$ were separated and ground into a powder. Approximately 500 mg of sample was obtained. Powder neutron diffraction data were collected at the Canadian Neutron Beam Centre using the C2 diffractometer at 20 K and 3.5 K. Wavelengths of 1.33 \AA\  and 2.37 \AA\  were used. The sample was contained in a thin-walled vanadium can and cooled in a closed-cycle refrigerator.

The single crystal diffraction on the $\beta$-NiNb$_2$O$_6$ polymorph was done on a sample that was from the same crystal as the crushed powder. The crystal was oriented initially using Laue back-scattering x-ray diffraction and then realigned with neutrons in situ at the start of the experiment. The single crystal neutron diffraction experiment was performed at the HB3A four-circle diffractometer at the High Flux Isotope Reactor (HFIR) at Oak Ridge National Laboratory (ORNL) in the USA~\cite{chakoumakos2011}. The sample was glued to the top of an aluminum rod and mounted in a closed-cycle helium-4 refrigerator with a temperature range of 4 to 450 K. The measurements used a monochromatic beam of neutrons with a wavelength of 1.546 \AA\, selected from the (220) Bragg reflection of a bent silicon monochromator. The scattered intensity was measured using an area detector and the data were analyzed using the Graffiti software package~\cite{lumsden2015} as well as the FullProf suite~\cite{rodriguezCarvajal1993}. More specifically, we generated appropriate data for the nuclear refinement at 20 K by first fitting the rocking curves through the measured peak positions to Gaussians, and then the extracted integrated intensities were corrected by the appropriate Lorentz factors to obtain the experimental $F^2$ values. A similar procedure was used for the magnetic refinements, although the magnetic signal was first isolated by subtracting the high-temperature 20 K dataset from the corresponding low-temperature 4 K dataset.

Muon spin relaxation ($\mu$SR)~\cite{schenck1985,lee1999,yaouanc2011} is a very sensitive technique that probes the internal magnetic field using the magnetic moment of the muon. These muons are implanted in a sample where they penetrate a few hundred $\mu$m, rapidly thermalize due to electrostatic interactions and finally settle at a Coulomb potential minimum in the material. Appropriate electronic vetoing removes any double counts due to cosmic ray muons or if multiple muons are generated and become incident on the sample within one observation window. A thin muon detector is placed at the entry to the sample chamber. The thin muon detector starts a clock registering the entry time of the muon. The muon will spontaneously decay into a positron and two neutrinos with a lifetime of 2.2 $\mu$s. In this decay, the positron is emitted preferentially in the direction of the muon spin at the time of decay. The positron will be detected in one of a number of scintilltors surrounding the sample and this will stop the clock and give a time from entry to decay. From the large ensemble of these events a histogram of positron counts in opposing detectors, $N_A$ and $N_B$ can be generated. Each positron counter will have an amplitude that fits the decay time of the muon with a signal representing the local field. By combining two opposing counters we can generate an asymmetry in the signal which is proportional to the spin polarization function: $A$~$=$~$\frac{N_A - N_B}{N_A + N_B}$. The LAMPF time-differential spectrometer with a helium-4 cryostat and a ultra low-background copper sample holder were used. This setup is capable of measuring in temperatures in the range of 2 to 300 K at fields from 0 to 4 kG applied longitudinally and allows for small transverse fields (up to $\sim$40 Gauss) to be applied using transverse Helmholtz coils. The time resolution for the data bins was set to 0.2 ns and the data collection window was set to 11 $\mu$s plus a 125 ns background measurement for subtraction.

The samples used in the $\mu$SR measurements were originally single crystals, but were crushed into a fine powder to remove orientation dependence in the ordered sample; Therefore we expect $\frac{2}{3}$ of the signal to be oscillatory in nature, which arises from components of the local magnetic moment that are perpendicular to the initial muon polarization. The remaining $\frac{1}{3}$ of the signal, arising from components of the local magnetic moment that are parallel to the initial muon spin direction, will cause no precession~\cite{kubo1981}. With zero applied field, when a sample is in a paramagnetic state with no static magnetism we will see a nearly time independent asymmetry, with deviations caused by the nuclear magnetic moments. In an ordered state, we will see an oscillating asymmetry caused by precession at a frequency that is proportional to the magnetic field at the stopping site multiplied by the muon gyromagnetic ratio, $\sim$135.5 MHz/T. The powder was secured in thin-walled Mylar in the sample holder.  All data was fit with the $\mu$SRfit data package~\cite{suter2012}.

\section{Results and Discussion}
\subsection{Magnetic Structure of the $\beta$-NiNb$_2$O$_6$ polymorph}

\begin{figure}[h]
	\centering
	\includegraphics[width=0.9\linewidth]{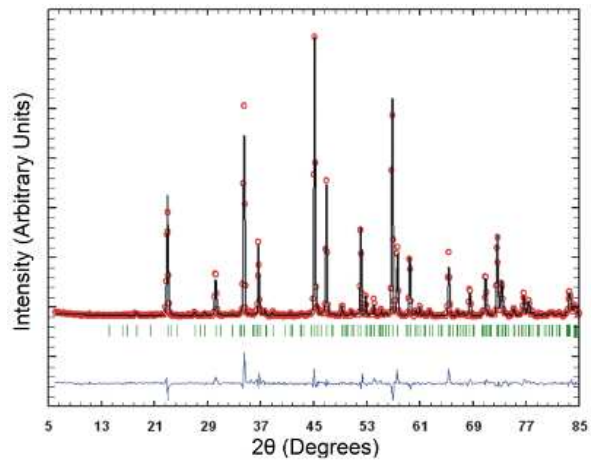}
	\caption{(Colour online) Rietveld refinement of neutron powder diffraction data at 20 K for $\beta$-NiNb$_2$O$_6$. The red circles are the data, the black line is the model, the blue line (below model) is the difference plot and the green tick marks locate the Bragg peaks. The wavelength is 1.33\AA.}
	\label{fig:SI1}
\end{figure}

We refined the powder data at 20 K, well above T$_N \sim$ 15 K, to check the sample quality and for any preferred orientation introduced by the grinding process. Our result shows good agreement with the single crystal model, allowing for a shrinking of the cell dimensions due to cooling. This is shown in Figure \ref{fig:SI1} and Table \ref{tbl:SI1}.

\begin{table}[h]
	\renewcommand{\arraystretch}{1.2}
	\begin{tabular}{|c|c|}
		\hline 
		a = 6.6851(4) \AA	& c = 9.0952(6) \AA \\ 
		\hline 
	\end{tabular} 
	\begin{tabular}{|c|c|c|c|c|}
		\hline 
		Atom	& x & y & z & B(\AA$^2$) \\ 
		\hline 
		Ni	& 0.5 & 0.5 & 0.5 & 0.3(1) \\ 
		\hline 
		Nb	& -0.014(1) & 0.4739(9) & 0.6660(9) & 0.5(1) \\ 
		\hline 
		O1	& 0.005(3) & 0.697(1) & 0.832(2) & 0.26(9) \\ 
		\hline 
		O2	& 0.008(5) & 0.695(1) & 0.501(2) & 0.26(9) \\ 
		\hline 
		O3	& 0.292(1) & 0.508(3) & 0.683(2) & 0.26(9) \\ 
		\hline 
	\end{tabular} 
	\begin{tabular}{|c|c|c|c|}
		\hline 
		R$_{wp}$	& 0.0461 & R$_{Bragg}$ & 0.0857\\ 
		\hline 
	\end{tabular} 
	\caption{Cell constants, atomic positions, displacement factors and agreement indicies for the refinement of powder neutron diffraction data for NiNb$_2$O$_6$ in the P4$_2$/n space group at 20 K.}
	\label{tbl:SI1}
\end{table}

\begin{figure}
	\centering
	\includegraphics[width=0.9\linewidth]{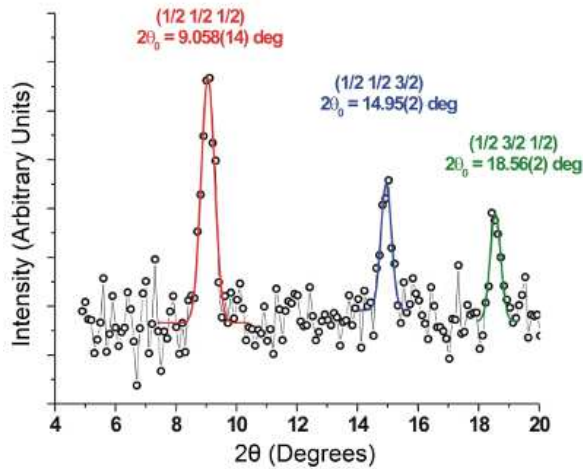}
	\caption{A difference plot (3.5 K - 20 K) showing three magnetic reflections indexed on $\vec{k}$ = ($\frac{1}{2},\frac{1}{2},\frac{1}{2}$). The neutron wavelength is 1.33\AA. The fits are to single Gaussians.}
	\label{fig:ninb2o6powderneutron}
\end{figure}

Data were then collected at 3.5 K to determine the magnetic structure. Three magnetic reflections could be identified and indexed with $\vec{k}$ = ($\frac{1}{2},\frac{1}{2},\frac{1}{2}$) as shown in Fig.~\ref{fig:ninb2o6powderneutron}.

\begin{figure}
	\centering
	\includegraphics[width=0.9\linewidth]{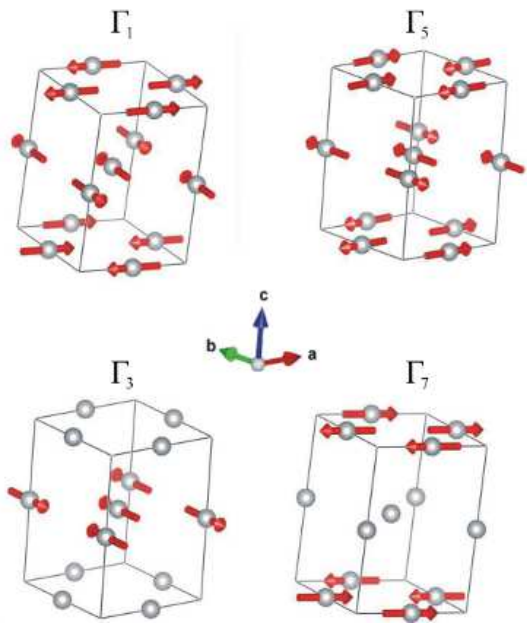}
	\caption{Four possible spin configurations for $\beta$-NiNb$_2$O$_6$ shown on the chemical unit cell. (Figure made using VESTA~\cite{momma2011})}
	\label{fig:ninb2o6magbasis}
\end{figure}

Attempts to solve the magnetic structure were aided by representation analysis using the program SARAh~\cite{wills2000}. For P4$_2$/n, $\vec{k}$ = $\left(\frac{1}{2},\frac{1}{2},\frac{1}{2}\right)$ and Ni$^{2+}$ at Wyckoff site 4e, there are four irreducible representations (IRs), $\Gamma_1$, $\Gamma_3$, $\Gamma_5$ and $\Gamma_7$, which provide potential basis vectors. The spin configurations consistent with these refinements are shown in Figure \ref{fig:ninb2o6magbasis} based on the chemical unit cell. 

%Attempts to solve the magnetic structure were aided by representation analysis using the program SARAh~\cite{wills2000}. For P4$_2$/n, $\vec{k}$ = $\left(\frac{1}{2},\frac{1}{2},\frac{1}{2}\right)$ and Ni$^{2+}$ at Wyckoff site 4e, there are four irreducible representations (IRs), $\Gamma_1$, $\Gamma_3$, $\Gamma_5$ and $\Gamma_7$, which provide potential basis vectors.  Refinements were attempted based  on these representations using the FullProf Suite~\cite{rodriguezCarvajal1993}. In an initial model, components of the Ni moment were permitted along the a-, b- and c-axes. Upon refinement, the components along a and c converged to values which were smaller than the errors, so these were set to zero. For $\Gamma_1$ and $\Gamma_5$ the Ni moment refined to 1.37(7) $\mu_B$ with R$_{mag}$~$=$~0.186. For $\Gamma_3$ and $\Gamma_7$ the Ni moment is 1.94(9) $\mu_B$ with essentially the same agreement factors. The spin configurations consistent with these refinements are shown in Figure \ref{fig:ninb2o6magbasis} based on the chemical unit cell. 

Note that the models fall into two sets that are intricately related and cannot be distinguished by unpolarized neutron scattering. All the Ni ions are ordered in the $\Gamma_1$ and $\Gamma_5$ models, while for $\Gamma_3$ and $\Gamma_7$ only Ni ions in every other layer are ordered. The $\Gamma_3$ and $\Gamma_7$, taken separately, seems physically unreasonable as well; there is no experimental evidence that half of the spins remain either paramagnetic or in a two-dimensional correlated state at 3.5 K. For example, there is no low temperature Curie-Weiss tail in the measured susceptibility~\cite{munsie2016} and the peak shape of the magnetic reflections, Fig.~\ref{fig:ninb2o6powderneutron}, shows no Warren-like feature. Additionally, since $\Gamma_3$ and $\Gamma_7$ form a co-representation, they can be combined to form a new symmetry-allowed model. Refinements on the powder using each of these models gave equally probable values for every model due to the small number of magnetic peaks obtained from the powder neutron scattering. We therefore collected single crystal neutron diffraction data to determine the magnetic structure.

\begin{figure}
	\centering
	\includegraphics[width=0.9\linewidth]{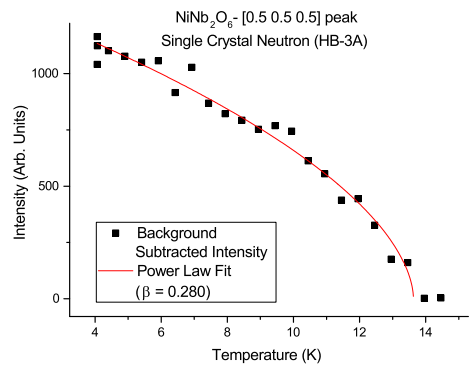}
	\caption{A plot of the T-dependence of the magnetic $\left(\frac{1}{2},\frac{1}{2},\frac{1}{2}\right)$ Bragg peak. A power law fit gives a value of the critical exponent $\beta$~$=$~0.28(3).}
	\label{fig:NeutronPowerLawFit}
\end{figure}

Two series of single crystal measurements were made. First, the intensity of the strongest magnetic peak, $\left(\frac{1}{2},\frac{1}{2},\frac{1}{2}\right)$ as determined from the powder neutron diffraction data, was measured from just above T$_N$ to 4 K and the results are shown in Fig.~\ref{fig:NeutronPowerLawFit}. The data below the critical temperature were then fit to a power law given by the expression:

\begin{equation}
\dfrac{I}{I_0} = A\left|\frac{T - T_C}{T_C}\right|^{2\beta}
\label{eqn:intensityBeta}
\end{equation}

where $\frac{I}{I_0}$ is the normalized intensity, $T$ is the temperature, $T_C$ is the transition (critical) temperature and $\beta$ is the magnetic critical exponent. The fitted values are $T_N$~$=$~13.6(2) K and $\beta$~$=$~0.28(3). Note that $\beta$ is at 1$\sigma$ of the expected value for the 3D Ising model (0.31), at 2$\sigma$ of the 3D XY model (0.345) and within 3$\sigma$ of the 3D Heisenberg model (0.36)~\cite{collins1989}. Nonetheless, the range of data available is too narrow and of insufficient point density near T$_C$ to determine an accurate value for $\beta$ and this must be left to further study.

The second set of scans performed was a survey of both magnetic and structural peaks at a temperature below (4 K) and above (20 K) the transition temperature. We first performed a single crystal nuclear refinement with the 20 K data, and the resulting parameters agreed well with those from our powder diffraction measurement. Magnetic refinements were then performed with a (4 K - 20 K) temperature difference dataset, which ensured that the magnetic scattering was isolated appropriately. We considered all candidate models discussed previously in the neutron powder diffraction section. 

\begin{figure}
	\begin{subfigure}{\columnwidth}
		\centering
		\includegraphics[width=.95\linewidth]{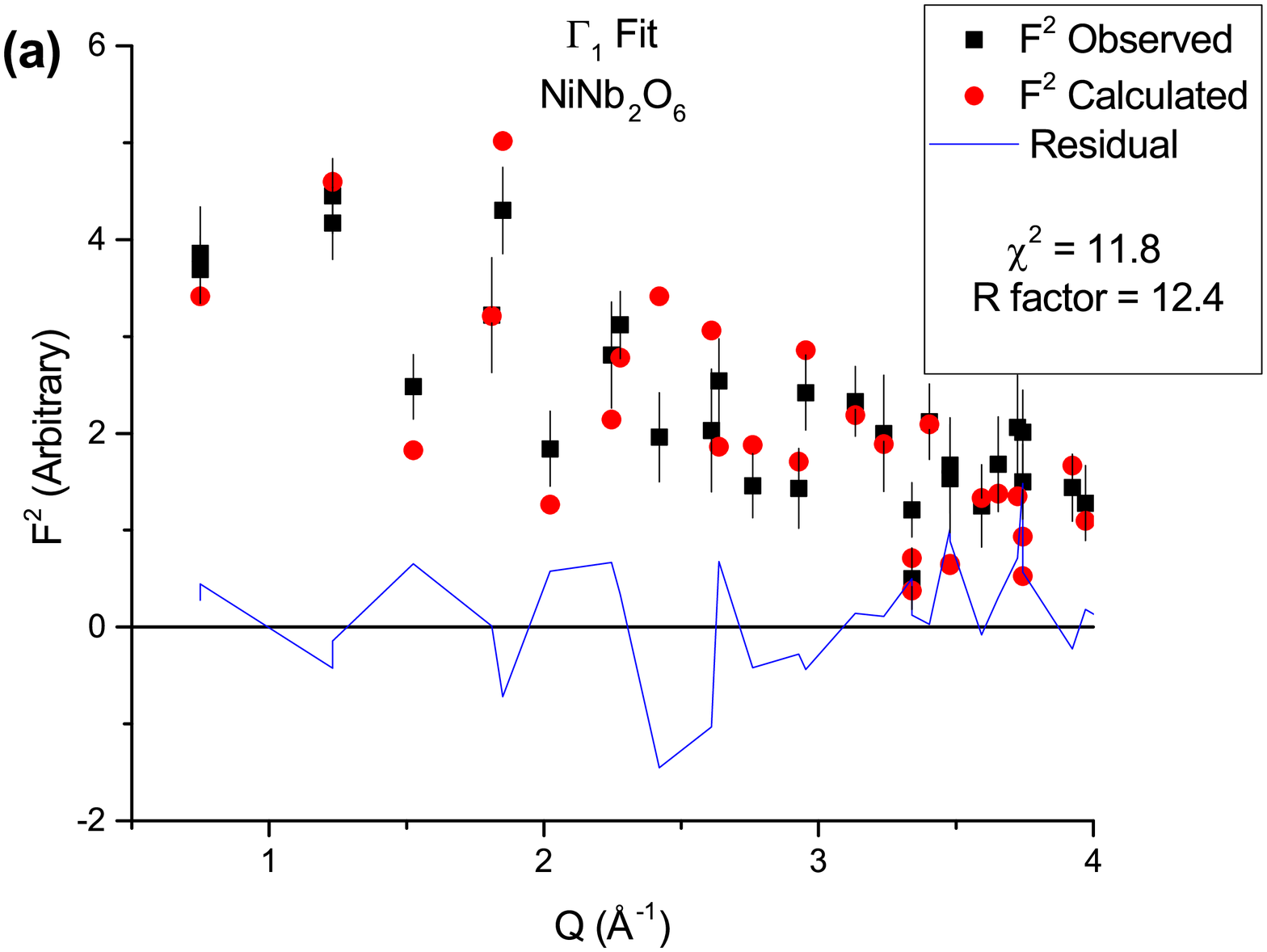}
		\caption{}
		\label{fig:gamma1f2}
	\end{subfigure}
	\begin{subfigure}{\columnwidth}
		\centering
		\includegraphics[width=.95\linewidth]{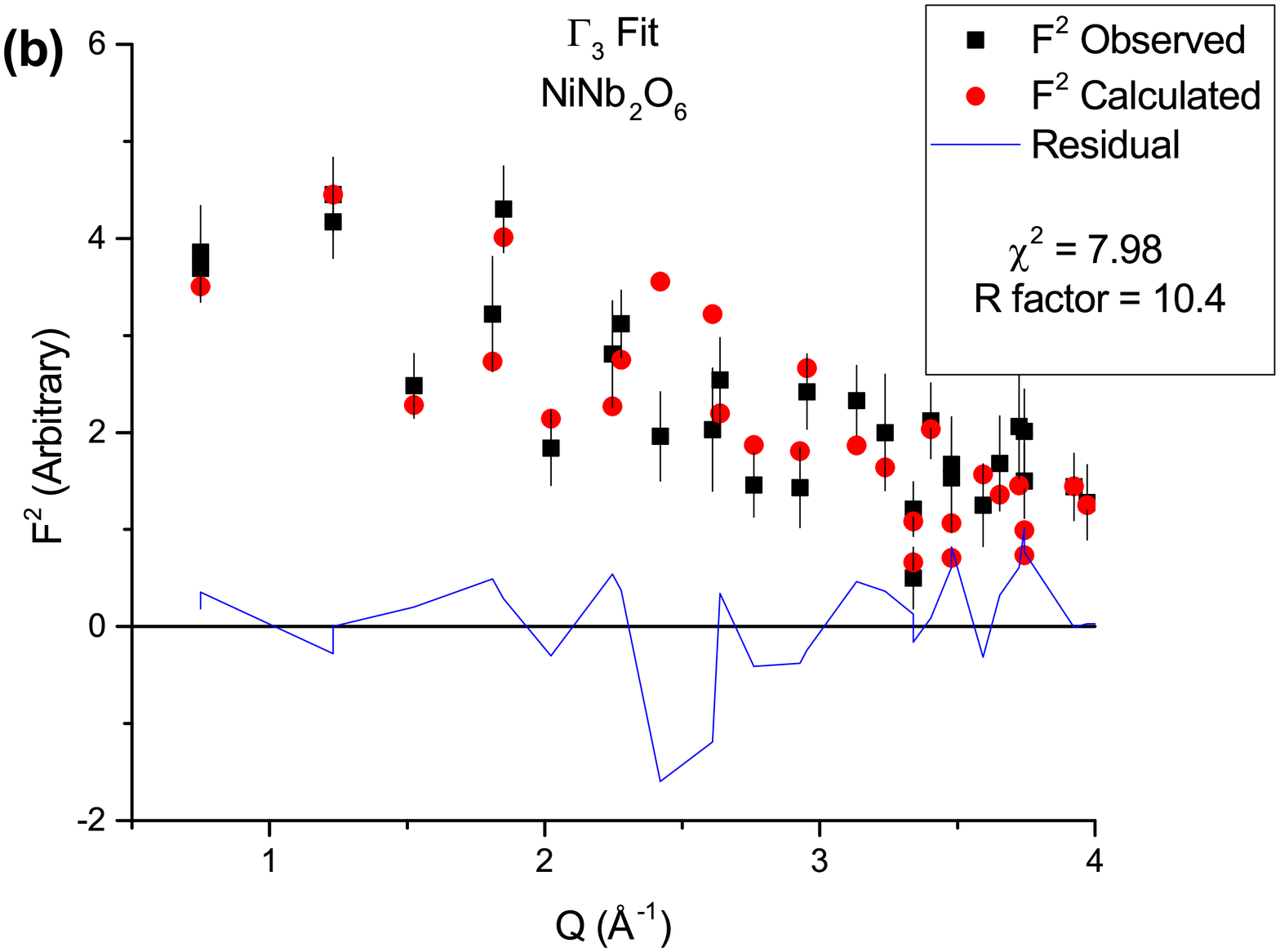}
		\caption{}
		\label{fig:gamma3f2}
	\end{subfigure}
	\begin{subfigure}{\columnwidth}
	\centering
	\includegraphics[width=.95\linewidth]{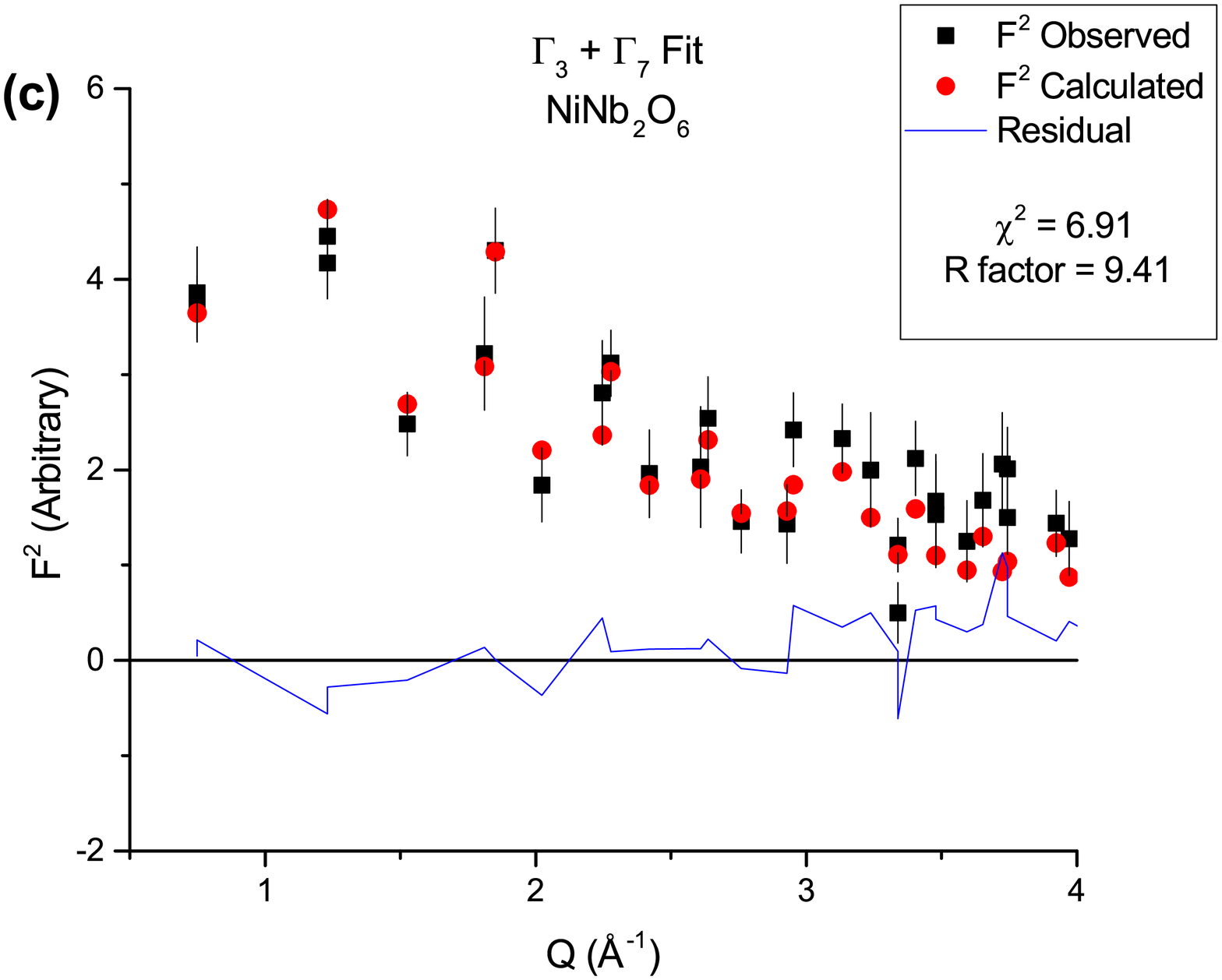}
	\caption{}
	\label{fig:gamma3gamma7f2}
\end{subfigure}
	\caption{A comparison of fits of the actual data from single crystal neutron scattering on NiNb$_2$O$_6$ to the expected scattering given a different magnetic basis vectors. The calculated and observed values of F$^2$, the residuals, $\chi^2$ and R-factor for (a) the $\Gamma_1$, (b) $\Gamma_3$, and (c) $\Gamma_3+\Gamma_7$ models are shown.}
	\label{fig:NeutronDataFitPlots}
\end{figure}

\begin{figure}
	\centering
	\includegraphics[width=0.7\linewidth]{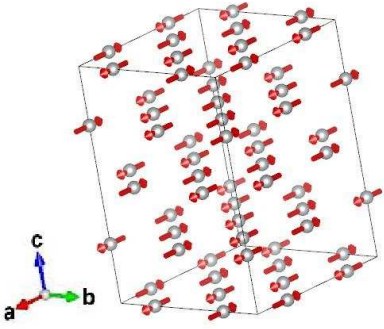}
	\caption{The full magnetic structure of NiNb$_2$O$_6$ in the $\Gamma_3 + \Gamma_7$ basis. (Figure made using VESTA~\cite{momma2011})}
	\label{fig:ninb2o6magstudio}
\end{figure}

The values of the Ni$^{2+}$ ordered moments are given in Table \ref{tbl:NiNb2O6SingleXtalMoments} for the $\Gamma_1$ and $\Gamma_3$ models, and the refinement quality is presented for these two cases in Fig.~\ref{fig:gamma1f2} and \ref{fig:gamma3f2}. Neither of these models fit the data well and the $\Gamma_3$ model yielded a Ni$^{2+}$ moment that is too large to be consistent with a simple spin-only $S$~$=$~1 picture. We also tried to fit the data with $\Gamma_3 + \Gamma_7$ model discussed previously. This model clearly provides the best fit to the data as indicated by the lowest $\chi^2$ and R-factor of the three refinements attempted; the refinement result for the $\Gamma_3 + \Gamma_7$ model is displayed in Fig.~\ref{fig:gamma3gamma7f2}. The Ni moments were originally unconstrained in this refinement, resulting in nearly collinear moments pointing along the a-axis in adjacent ab-layers. The moments were then constrained to be collinear in adjacent layers and point exactly along the a-axis, which changed the goodness of fit negligibly. For this reason, the magnetic structure of the constrained refinement is considered to be the final solution; the ordered Ni$^{2+}$ moment obtained in this case is shown in Table~\ref{tbl:NiNb2O6SingleXtalMoments} and a schematic of the structure is presented in Fig.~\ref{fig:ninb2o6magstudio}. The determined magnetic structure for $\beta$-NiNb$_2$O$_6$ is consistent with the exchange interaction hierarchy expected from crystal structure considerations. As explained previously, the strongest exchange interaction likely corresponds to the in-plane 180$^\circ$ Ni--O--O--Ni pathway, which should be strongly antiferromagnetic. We note that this strong exchange pathway alternates along the a- and b-axes in adjacent ab-plane layers of Ni$^{2+}$. Furthermore, all other superexchange interactions are mediated by two or three ions with less favourable geometries for strong orbital overlap, and therefore one expects $\beta$-NiNb$_2$O$_6$ to be best described by a series of coupled antiferromagnetic Ni$^{2+}$ chains~\cite{law2014} that ultimately achieve long-range order due to weak interchain interactions.

\begin{table}
	\begin{tabular}{|c|c|c|c|}
		\hline 
		Moment	& $\Gamma_1$ moments & $\Gamma_3$ moments & $\Gamma_3 + \Gamma_7$ moments \\ 
		\hline 
		$\mu_a$	& 0.44(6) & 2.15(3) & 1.60(1)\\ 
		\hline 
		$\mu_b$	& 1.67(2) & 0.41(3) & 0\\ 
		\hline 
		$\mu_c$	& 0.16(3) & 1.00(4) & 0\\ 
		\hline 
		Total & 1.73(5) & 2.41(6) & 1.60(1) \\ 
		\hline
		
	\end{tabular} 
	\caption{The values of the refined moments from single crystal neutron diffraction data for various models.}
	\label{tbl:NiNb2O6SingleXtalMoments}
\end{table}

\begin{figure}
	\centering
	\includegraphics[width=\linewidth]{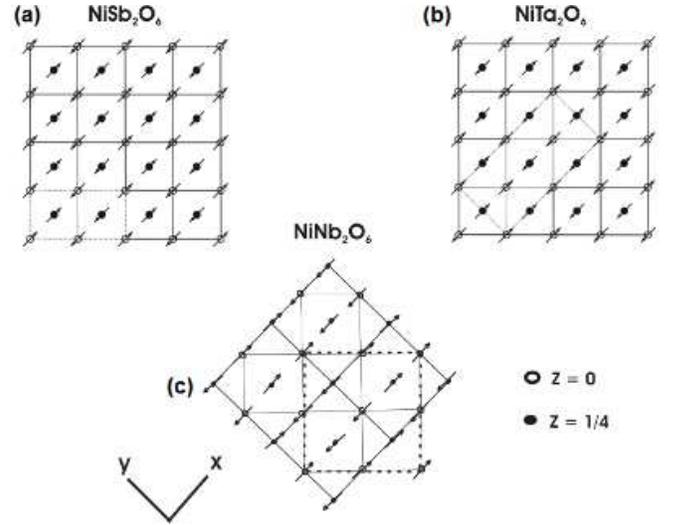}
	\caption{A comparison of the magnetic structures for (a) NiSb$_2$O$_6$~\cite{ehrenberg1998}, (b) NiTa$_2$O$_6$~\cite{ehrenberg1998} and (c) $\beta$-NiNb$_2$O$_6$ projected along the c-axis and with two adjacent layers shown. In (c) the larger chemical cell is indicated by the darker solid lines and the smaller TR-like chemical cell by the lighter solid lines. In all three diagrams the dashed lines indicate the magnetic unit cell.\footnote{Images (a) and (b) reprinted from Journal of Magnetism and Magnetic Materials, 184, Ehrenberg, H; Wltschek, G; Rodriguez-Carvajal, J; and Vogt, T, Magnetic Structures of the Tri-Rutiles NiTa$_2$O$_6$ and NiSb$_2$O$_6$, 111--115, 1998, with permission from Elsevier}}
	
	\label{fig:magcomparison}
\end{figure}

When comparing the magnetic structure for NiNb$_2$O$_6$ with those for NiSb$_2$O$_6$~\cite{ehrenberg1998} and NiTa$_2$O$_6$~\cite{ehrenberg1998} in Fig.~\ref{fig:magcomparison}, there are interesting similarities and differences. The magnetic structure of all three systems  satisfies the dominant AF exchange interaction described above. More specifically, all three materials consist of moments oriented along the [110] direction in the TR or TR-like cell that form AF-coupled chains running along the same direction, consistent with Fig.~\ref{fig:spacegroupninb2o63} and DFT calculations for NiTa$_2$O$_6$~\cite{law2014}. Differences arise in the coupling of these AF chains to adjacent chains both within and between layers, which emphasizes the importance of the interchain interactions on determining the precise magnetic structure. In fact, the magnetic structures and the magnetic unit cells expressed in terms of the parent TR chemical cell are different for all three phases. For NiSb$_2$O$_6$ a$_M$ = 2a$_{TR}$, b$_M$ = b$_{TR}$  and c$_M$ = 2c$_{TR}$ giving V$_M$ = 4V$_{TR}$. For NiTa$_2$O$_6$ a$_M$ = 2$^{\frac{1}{2}}$a$_{TR}$, bM = 2(2$^{\frac{1}{2}}$)b$_{TR}$ and c$_M$ = 2c$_{TR}$, giving V$_M$ = 8V$_{TR}$ while for NiNb$_2$O$_6$ a$_M$ = 2a$_{TR}$, b$_M$ = 2b$_{TR}$ and c$_M$ = 2c$_{TR}$  also yielding V$_M$ = 8V$_{TR}$. This is a remarkable result, given the close structural similarities among the three materials. The exact position of the ligands relative to the metal ions seems to serve as a very sensitive tuning parameter for the weak interchain interactions, which leads to the diverse magnetic ground states observed for this family of materials.

\subsection{Zero Field $\mu$SR on the $\beta$-NiNb$_2$O$_6$ polymorph}

\begin{figure*}
\begin{subfigure}{.5\textwidth}
  \centering
  \includegraphics[width=\linewidth]{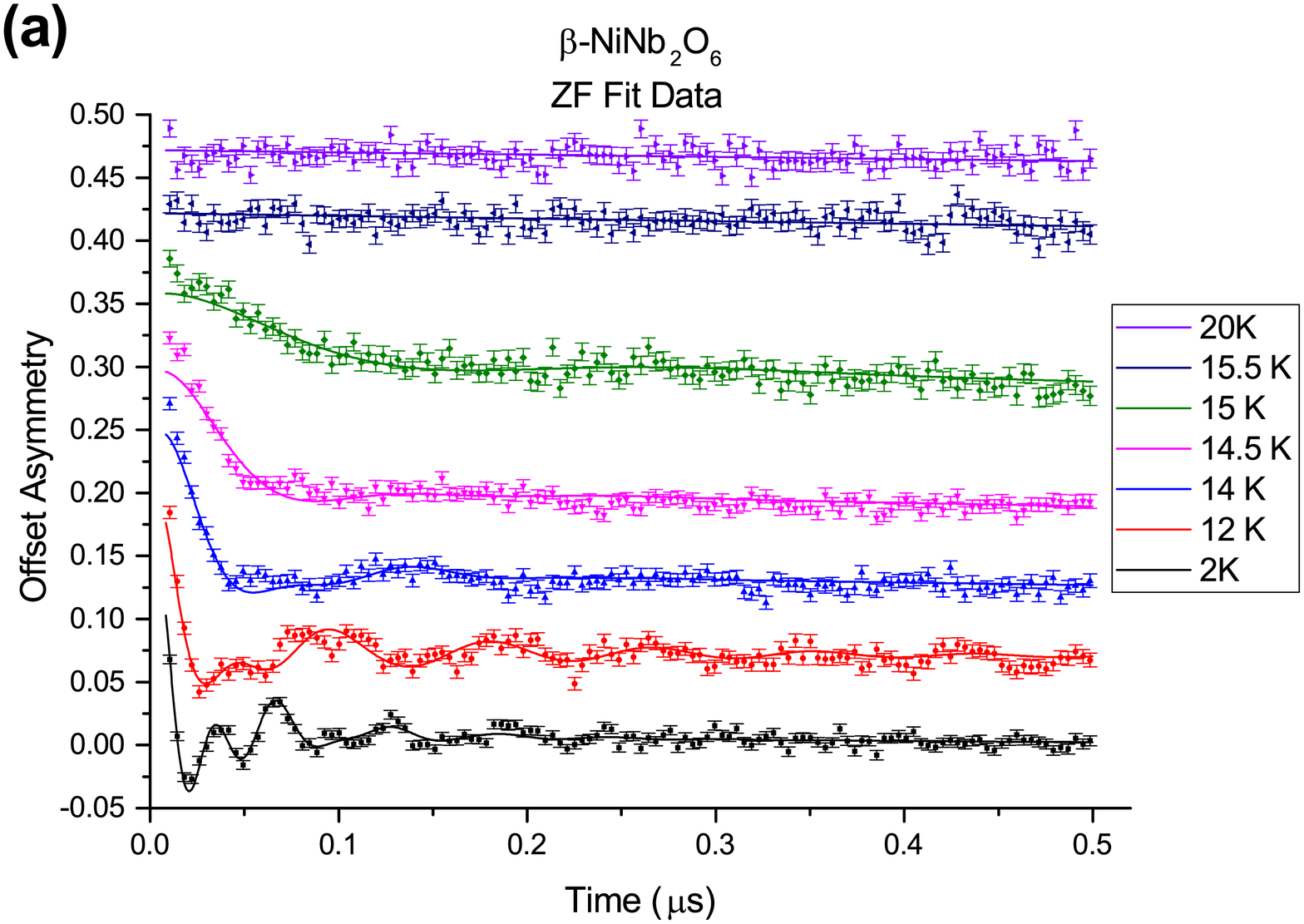}
  \caption{}
  \label{fig:ZFAsyFitNew}
\end{subfigure}%
\begin{subfigure}{.5\textwidth}
  \centering
  \includegraphics[width=\linewidth]{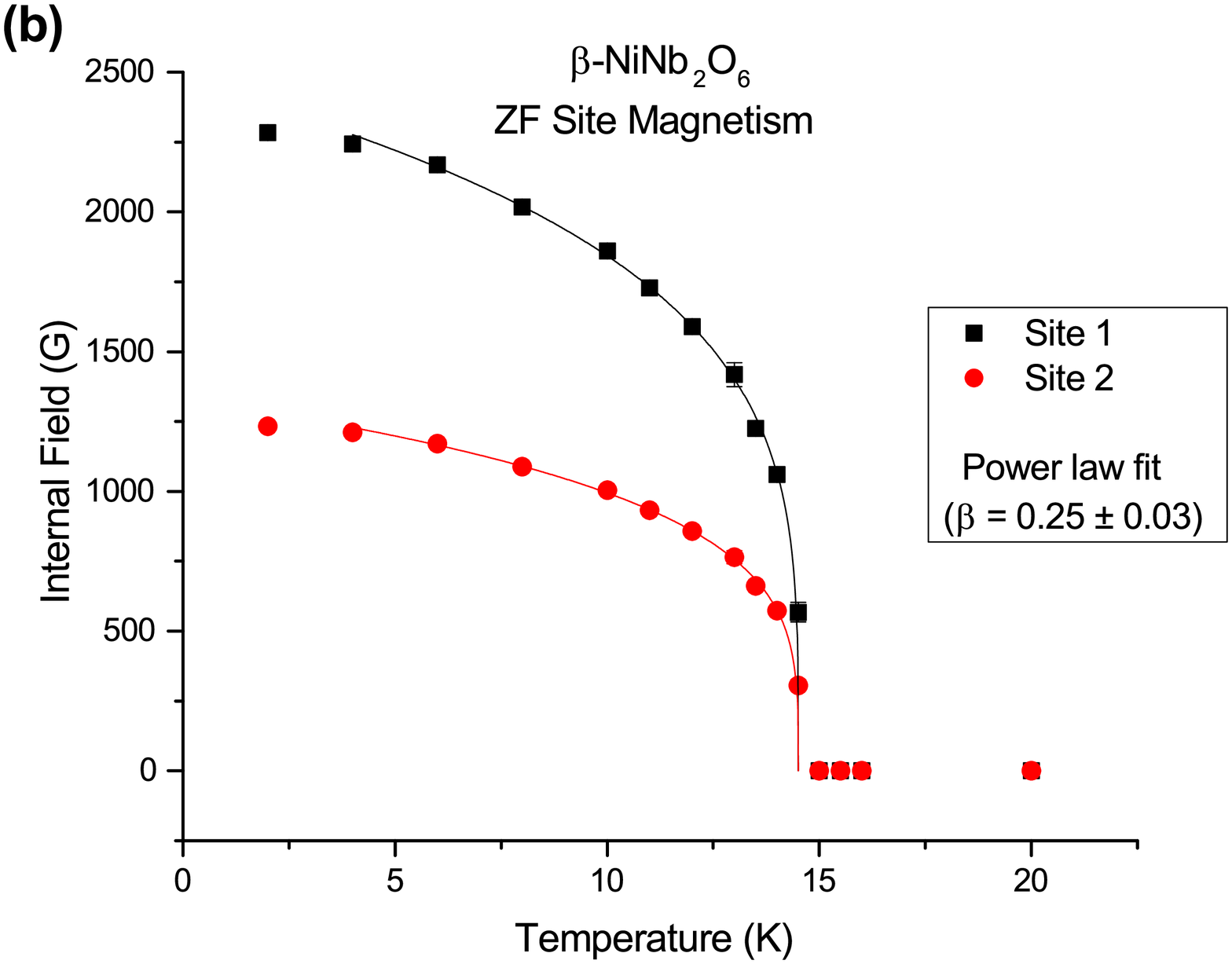}
  \caption{}
  \label{fig:ZFFieldFitNew}
\end{subfigure}
\caption{Selected data and internal field fits from zero field $\mu$SR measurements performed on the $\beta$-NiNb$_2$O$_6$ polymorph. Panel (a) (Colour online) shows fits of $\mu$SR data from 2 to 20 K for $\beta$-NiNb$_2$O$_6$. Each asymmetry plot is offset by +0.05 for clarity. There is a distinct reduction in the local field with increasing temperature, followed by a transition to the paramagnetic state above 15 K. In panel (b), we plot the internal magnetic field strength at two different muon sites in $\beta$-NiNb$_2$O$_6$.}
\label{fig:ZFMuSRNew}
\end{figure*}

Zero field $\mu$SR measurements of $\beta$-NiNb$_2$O$_6$ are presented in Fig.~\ref{fig:ZFAsyFitNew}. Asymmetry spectra at several temperatures with superimposed fits are shown. Each of the spectra are offset by a value of 0.05 per temperature step for clarity. At 2 K, it is evident that in the first 0.2 $\mu$s there is strong oscillation in the asymmetry, indicative of long range order in the material. A fast Fourier transform showed that there are likely two oscillating components, so the data was fit to the 3-component form given by the following expression: 

\begin{equation}
\begin{split}
A\left(t\right) = &A_1\cos\left(\gamma_\mu Bt + \phi \right)e^{-\lambda_{1} t} + \\&A_2\cos\left(\gamma_\mu \delta Bt + \phi \right)e^{-\frac{\left(\sigma_{2} t\right)^2}{2}} + A_3e^{-\lambda_{3} t}
\end{split}
\label{eqn:ZFNewEqn}
\end{equation}

Fitting this form, the value of $\alpha$, the phase $\phi$ and the total asymmetry (A$_1$ + A$_2$ + A$_3$) were set to a constant determined by fitting a weak transverse field run. The scaling factor between the two fields (B) was fit globally and found to be 0.540(3). $\gamma_\mu$ is the muon gyromagnetic ratio.  $\lambda_n$ are the relaxation rates in $\mu$s$^{-1}$ of the n$^{\text{th}}$ term. The first two terms are relaxing, oscillating components with an exponential (first term) and Gaussian (second term) envelope. The final term is a non-precessing component, which is fitting the long time tail of the signal, typical of relaxation due to fluctuations of the local field. Each individual asymmetry was fit as a free parameter.

The values of the local fields at the two muon stopping sites, as determined by fits to the data, are plotted in figure \ref{fig:ZFFieldFitNew}. The extracted temperature dependence of the two local fields were fit simultaneously to a power law model, giving a value for the critical exponent $\beta = 0.25(2)$. This was fit using: 

\begin{equation}
F = A\left|\dfrac{T-T_C}{T_C}\right|^\beta
\label{eqn:fieldBeta}
\end{equation}

where $F$ is the internal field, $T$ is the temperature, $T_C$ is the transition (critical) temperature and $\beta$ is the critical exponent for the order parameter. A value of $T_N$~$=$~14.5(3) K was determined from the fit, which agrees well with previous measurements \cite{munsie2016}. The fit also yielded a value for the critical exponent $\beta = 0.25(3)$, which is lower than typical values for 3D universality classes. In this case all three classes are at least 3$\sigma$ away from the fit value. This implies that reduced dimensionality is important in this material and is consistent with both previous DFT work~\cite{law2014} and the magnetic structure determined by neutron diffraction above.

\begin{figure*}
\begin{subfigure}{.5\textwidth}
  \centering
  \includegraphics[width=\linewidth]{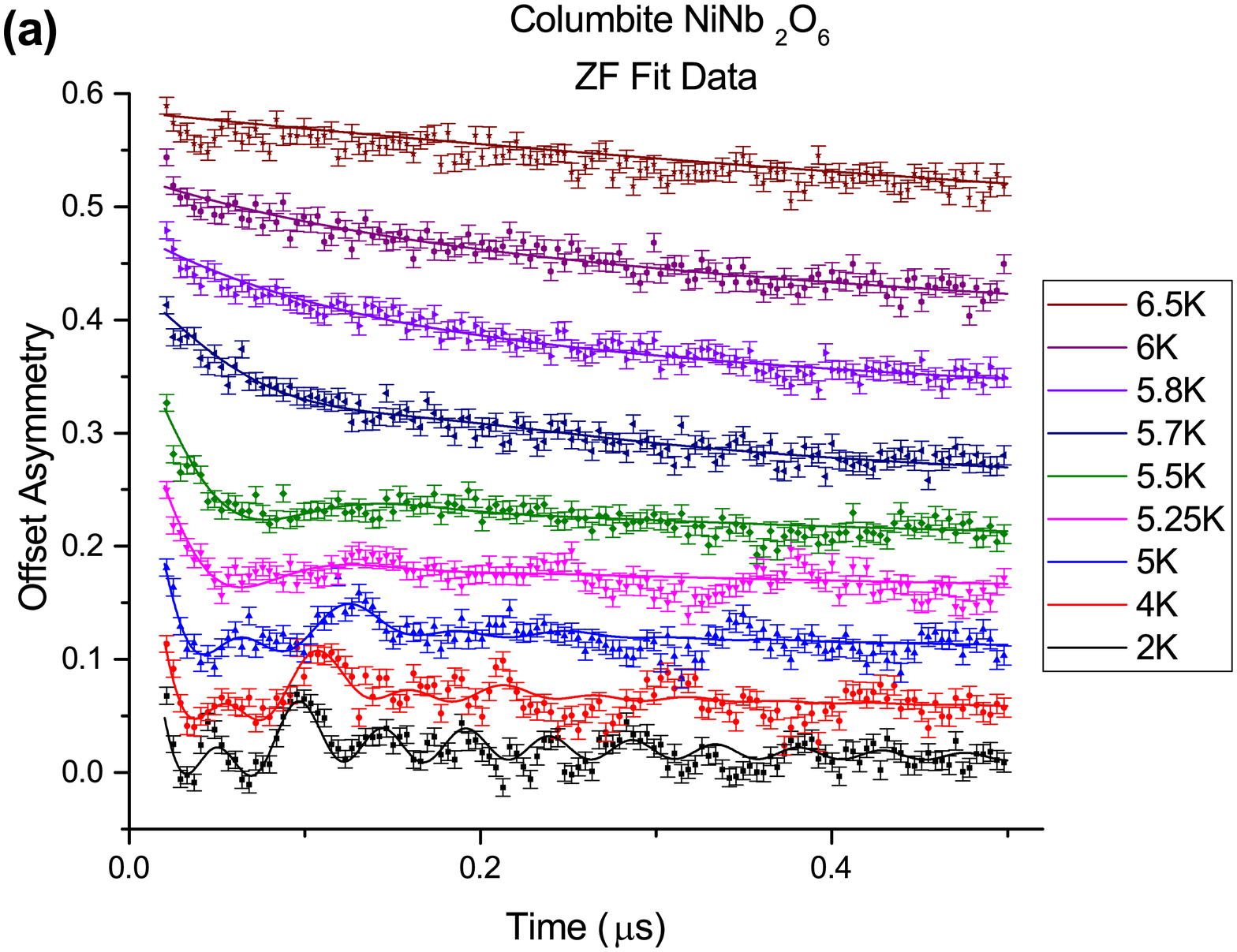}
  \caption{}
  \label{fig:ZFAsyFitCol}
\end{subfigure}%
\begin{subfigure}{.5\textwidth}
  \centering
  \includegraphics[width=\linewidth]{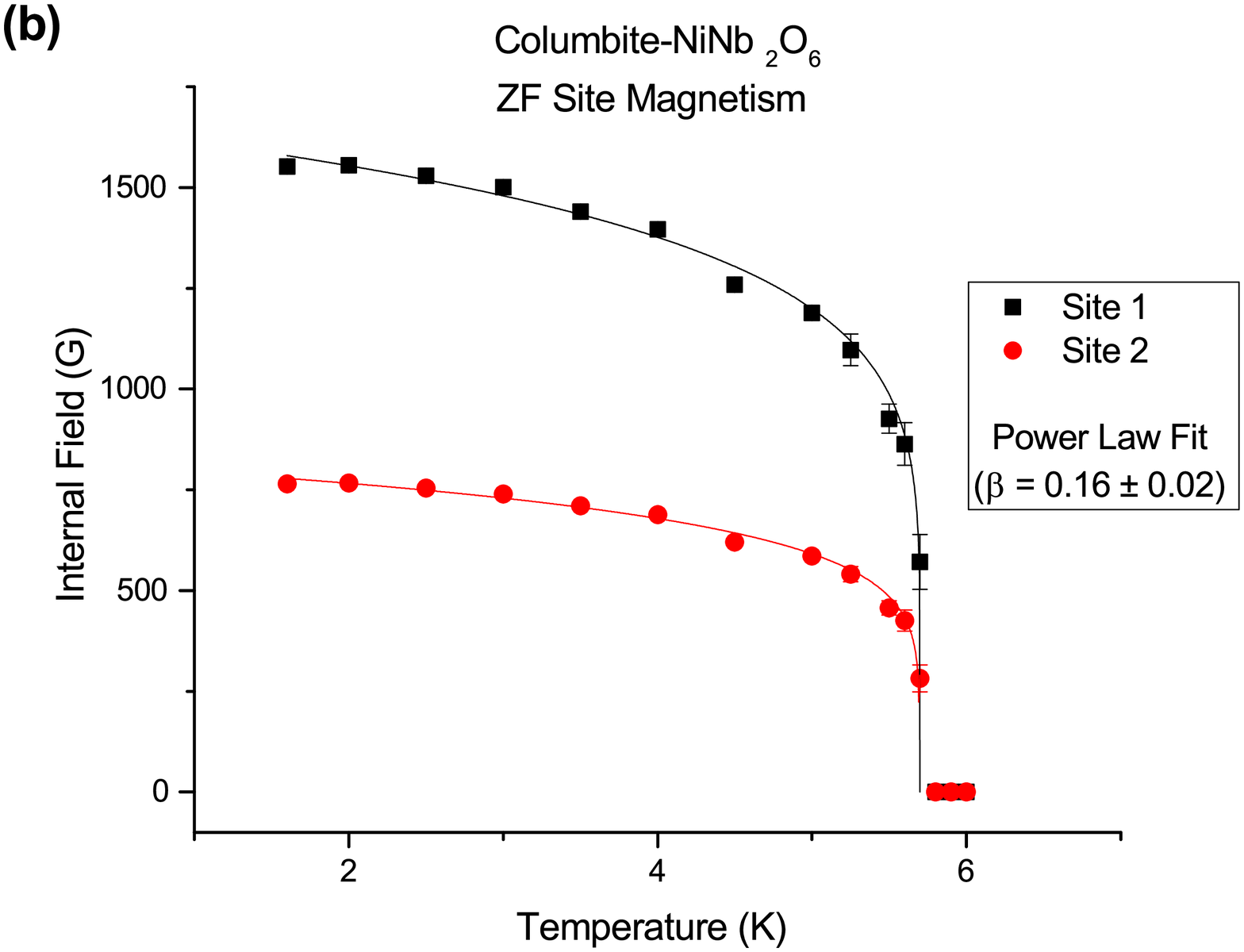}
  \caption{}
  \label{fig:ZFFieldFitCol}
\end{subfigure}

\caption{Selected data and internal field fits from zero field $\mu$SR measurements performed on the columbite polymorph of NiNb$_2$O$_6$. In panel (a) (Colour online) we show fits of $\mu$SR data from 2 to 20 K for the columbite polymorph of NiNb$_2$O$_6$. Each asymmetry plot is offset by +0.05 for clarity. There is a distinct decrease in the local field with increasing temperature, followed by a transition to the paramagnetic state above 5.7 K. In panel (b) we plot the internal magnetic field strength at two different muon sites in the columbite polymorph of nickel niobate.}
\label{fig:ZFMuSRCol}
\end{figure*}

\subsection{Zero Field $\mu$SR on the columbite polymorph}

Similar $\mu$SR measurements were performed on the crushed single crystal powder sample of the columbite polymorph of nickel niobate. The results from the temperature sweep with zero field applied are shown in figure \ref{fig:ZFMuSRCol}. From a fast Fourier transform, it was again evident that there were two primary components. As in the case for the $\beta$-NiNb$_2$O$_6$ polymorph, there was also a relaxing tail. The function used to fit this data was:

\begin{equation}
\begin{split}
A\left(t\right) = &A_1\cos\left(\gamma_\mu Bt + \phi \right)e^{-\lambda_{1} t} + \\&A_2\cos\left(\gamma_\mu \delta Bt + \phi \right)e^{-\lambda_{2} t} + A_3e^{-\lambda_{3} t}
\end{split}
\label{eqn:ZFColEqn}
\end{equation}

The difference between this function and Eqn. \ref{eqn:ZFNewEqn} is that both oscillating components have exponential front ends. Here we found the ratio, $\delta$ to be 0.493(4).

In Fig.~\ref{fig:ZFAsyFitCol}, it is evident that there is a very slow decrease in the oscillation frequency with increasing temperature between 2 and 5 K. This trend is followed by a very large drop in frequency above 5 K, and the oscillations disappear completely between 5.25 and 6 K. Finally, by 10 K (not shown) there is only a very weak relaxation of the signal remaining.

Figure \ref{fig:ZFFieldFitCol} plots the internal field strength against temperature. A power law fit was performed using Eqn.~\ref{eqn:fieldBeta}. A transition temperature of T$_N$~$=$~5.7(3) K was found from the fit, which is in good agreement with previous work\cite{yaeger1977,heid1996}. The fit also revealed a critical exponent $\beta$~$=$~0.16(2), which is even lower than the value found for the $\beta$-polymorph and therefore further away from typical 3D universality class exponents. This result emphasizes that columbite NiNb$_2$O$_6$ is a low-dimensional magnetic system and likely can be described by weakly-coupled $S$~$=$~1 spin chains; therefore it is the true $S$~$=$~1 cousin to the interesting $S_{eff}$~$=$~$\frac{1}{2}$ ferromagnetic Ising chain system CoNb$_2$O$_6$.

\section{Conclusion}
We have presented both neutron diffraction and $\mu$SR data on the $\beta$-polymorph of NiNb$_2$O$_6$ as well as $\mu$SR data on the columbite polymorph. Combined powder and single crystal neutron diffraction data allowed us to solve the magnetic structure unambiguously for $\beta$-NiNb$_2$O$_6$. We find a collinear arrangement of Ni$^{2+}$ moments, which is consistent with a dominant in-plane AF exchange path that alternates along the a and b axes in adjacent Ni$^{2+}$ layers. $\beta$-NiNb$_2$O$_6$ is best described as an $S$~$=$~1 chain system with significant interchain interactions, ultimately leading to long-range magnetic order at $T_N$~$=$~15 K. This picture is supported by $\mu$SR measurements, which find a critical exponent for the order parameter $\beta$~$=$~0.25(3).  $\mu$SR was also used to determine $\beta$~$=$~0.16(2) in the case of columbite polymorph. This exceptionally small value indicates that low dimensionality is an intrinsic property of the columbite polymorph, which is consistent with expectations for a weakly-coupled $S$~$=$~1 chain system. Inelastic neutron scattering measurements will be indispensable for determining the coupling strength between the Ni$^{2+}$ chains in these two polymorphs, and it will be extremely interesting to compare the inelastic neutron scattering spectrum of columbite NiNb$_2$O$_6$ to the cobalt analogue. \newline

\section{Acknowledgements}

Partial funding for this research came from an Ontario Graduate Scholarship (OGS) award and National Science and Engineering Research Council (NSERC) Grants. Research at the High Flux Isotope Reactor at the Oak Ridge National Laboratory was sponsored by the Scientific User Facilities Division, Office of Basic Energy Sciences, US Department of Energy. Research at the Canadian Neutron Beam Centre was supported by the Canadian Nuclear Laboratories, Chalk River, Canada.

We would like to thank A.M. Hallas for assistance with the preliminary work as well as useful discussions throughout the data collection and writing process. We would like to thank the staff at TRIUMF National Laboratory, specifically Dr. G. Morris and Dr. B. Hitti for their assistance, guidance and support throughout the $\mu$SR measurements.\newline

\end{document}